\documentstyle[aps,preprint]{revtex}
\newcommand{\be}{\begin{equation}}
\newcommand{\ee}{\end{equation}}
\newcommand{\beq}{\begin{eqnarray}}
\newcommand{\eeq}{\end{eqnarray}}
\newcommand{\cM}{\mbox{${\cal M}$}}
\newcommand{\lie}[1]{\mbox{$:\! #1 \!:$}}
\tightenlines
\begin{document}
\draft
\title{Symplectic Integration of Hamiltonian Systems using Polynomial Maps}
\author{Govindan Rangarajan$^{*}$}
\address{Department of Mathematics and Center for Theoretical
Studies, Indian Institute of Science, Bangalore 560 012, India}
\maketitle

\begin{abstract}
In order to perform numerical studies of long-term stability in
nonlinear Hamiltonian systems, one needs a numerical integration
algorithm which is symplectic. Further, this algorithm should be
fast and accurate. In this paper, we propose such a symplectic
integration algorithm using polynomial map refactorization of the
symplectic map representing the Hamiltonian system. This method
should be particularly useful in long-term stability studies of
particle storage rings in accelerators.
\end{abstract}

\vskip 12pt 
\newpage

\section{Introduction}

Long-term single particle stability studies of particle storage rings play an important
role in the design of accelerators \cite{BNL}. These storage rings are generally
described by nonlinear, nonintegrable Hamiltonians. Therefore analytical results
on long-term stability of particle motion in such storage rings are difficult to obtain.
By default, numerical integration of particle trajectories is the primary tool
used to explore the dynamics of these systems. However,
standard numerical integration algorithms can not be used since they are not
symplectic \cite{Dragt}. This violation of the symplectic condition can lead to spurious
chaotic or dissipative behavior. Numerical integration algorithms which
satisfy the symplectic condition are called symplectic integration algorithms
\cite{channel}.

Several symplectic integration algorithms have been proposed in the literature
\cite{list}. Some of these directly use the Hamiltonian whereas others use
the symplectic map \cite{Dragt} representing either the entire storage ring
(in which case one obtains the so-called one-turn map) or major segments of
the ring. For complicated systems like the Large Hadron Collider which has
thousands of elements, using individual Hamiltonians for each element can drastically
slow down the integration process. One the other hand, the map based approach
is very fast in such cases \cite{yan}. Further, if nonlinearities in the symplectic
map are too ``large'', one can use scaling and squaring techniques \cite{dragtprl}
to alleviate the problem.

One class of the map-based methods uses jolt
factorization \cite{irwin,list2}. But there are still unanswered questions on how to
best choose the underlying group and elements in the group \cite{etienne}.
Further, some of these methods \cite{list2} can be quite difficult to generalize to
higher dimensions. Another class of methods uses solvable maps \cite{grsolve}
or monomial maps \cite{gjaja}. Even though they are fairly straightforward
to generalize to higher dimensions, they tend to introduce spurious poles and branch
points not present in the original map \cite{etienne}.

In this paper, we propose a new symplectic integration method
where the symplectic map is refactorized using ``polynomial maps''
(maps whose action on phase space variables gives rise to
polynomials). This does not introduce spurious poles and branch
points. Moreover, it is easy to generalize to higher dimensions.
In this letter, we restrict ourselves to maps in six dimensional
phase space which are appropriate for single particle stability
studies. We show that the method gives good results. Further,
since it is map-based, it is also very fast.

\section{Preliminaries}

We start by representing a Hamiltonian system by a symplectic map \cite{Dragt}.
For simplicity we restrict ourselves to a six dimensional phase space.
Let us denote the collection of
six phase-space variables $q_i$, $p_i$ ($i=1,2,3)$ by the symbol $z$:
\be
z = (q_1,q_2,q_3,p_1,p_2,p_3).
\ee
The Lie operator \cite{Dragt} corresponding to a phase-space function $f(z)$ is denoted
by $\lie{f(z)}$.  It is defined by its action on a phase-space function $g(z)$
as shown below
\be
\lie{f(z)} g(z) = [f(z),g(z)].
\label{1a}\ee
Here $[f(z),g(z)]$ denotes the usual Poisson bracket of the functions $f(z)$
and $g(z)$.  Next, we define the exponential of a Lie operator.  It is
called a Lie transformation \cite{Dragt} and is given as follows:
\be
e^{:f(z):} = \sum_{n=0}^{\infty} \frac{\lie{f(z)}^n}{n!}.
\label{1b}\ee

The effect of a Hamiltonian system on a particle can be formally expressed
as the action of a map \cM\ that takes the particle from its
initial state $z^{in}$ to its final state $z^{fin}$
\be
z^{fin} = \cM\, z^{in}.
\label{I.15}\ee
It can be shown that ${\cal M}$ is a symplectic map \cite{Dragt}.
Consider its Jacobian matrix which we denote by $M$.
Symplectic maps are maps whose
Jacobian matrices $M$ satisfy the following `symplectic condition'
\be
\widetilde{M}JM = J,
\label{1e}\ee
where $\tilde{M}$ is the transpose of $M$ and $J$ is the fundamental
symplectic matrix.

Using the Dragt-Finn factorization theorem\cite{Finn,Dragt}, the symplectic map
${\cal M}$ can be factorized as shown below:
\be
{\cal M} = {\hat M} e^{:f_3:}\,e^{:f_4:}\, \ldots e^{:f_n:} \ldots \ .
\label{2}\ee
Here $f_n(z)$
denotes a homogeneous polynomial (in $z$) of degree $n$ uniquely
determined by the factorization theorem.
Further ${\hat M}$ gives the linear part of the map and hence has an
equivalent representation in terms of the Jacobian matrix $M$ of the
map ${\cal M}$ \cite{Dragt}:
\be
{\hat M} z_i = M_{ij} z_j = (Mz)_i.
\label{3}\ee
The infinite product of Lie transformations $\exp(\lie{f_n})$ $(n=3,4, \ldots$)
in Eq. (\ref{2}) represents the nonlinear part of ${\cal M}$.

Using the above procedure, one can represent each element in the storage ring
by a symplectic map. By concatenating \cite{Dragt} these maps together, we obtain the
so-called `one-turn' map representing the entire storage ring. This concatenation
is made possible by the Campbell-Baker-Hausdorff (CBH) theorem \cite{Cornwell}.
The one-turn map
gives the final state $z^{(1)}$ of a particle after one turn around the ring as a
function of its initial state $z^{(0)}$:
\be
z^{(1)} = {\cal M} z^{(0)}.
\ee
To obtain the state of a particle after $N$ turns, one has to merely
iterate the above mapping $N$ times i.e.
\be
z^{(N)} = {\cal M}^n z^{(0)}.
\label{rec} \ee
Since ${\cal M}$ is explicitly symplectic, this gives a symplectic integration
algorithm. Further, since the entire ring can be represented by a single (or
at most a few) symplectic map(s), numerical integration of particle trajectories using
symplectic maps is very fast.

\section{Symplectic Integration Using Polynomial Maps}

It is obvious that one can not use ${\cal M}$ in the form given in
Eq. (\ref{2}) for any practical computations.  It involves an
infinite number of Lie transformations.  Therefore, we have to truncate
${\cal M}$ by stopping after a finite number of Lie transformations:
\be
{\cal M} \approx {\hat M} e^{:f_3:}\,e^{:f_4:}\, \ldots e^{:f_P:}.
\label{4}\ee
However, each exponential $e^{:f_n:}$
in ${\cal M}$ still contains an infinite number of terms in its Taylor
series expansion.  One possible solution is to truncate the Taylor
series generated by the Lie transformations to order $P$. But this
violates the symplectic condition.

We get around the above problem by refactorizing ${\cal M}$ in terms
of simpler symplectic maps which can be evaluated exactly without truncation.
We use `polynomial maps' which give rise to polynomials when acting on the
phase space variables. This avoids the problem of spurious poles and
branch points present in generating function methods \cite{etienne},
solvable map \cite{grsolve} and monomial map \cite{gjaja} refactorizations.
To determine which symplectic maps give rise to polynomial mappings,
consider $\exp(:h(z):)$ where $h(z)$ is a polynomial. Since all
Lie transformations are symplectic maps \cite{Dragt}, this is a
symplectic map. Its action on phase space variables is equivalent
to solving the Hamilton's equations of motion from time $t=0$ to
$t=-1$ using $h(z)$ as the Hamiltonian. For example, consider
the action of $\exp(:q_1^3:)$ on $q_1$, $p_1$ in a two dimensional
phase space. We first solve the following Hamilton's equations of
motion:
\beq \label{Heqns}
\frac{dq_1}{dt} & = & \frac{\partial h}{\partial p_1}, \nonumber \\
\frac{dp_1}{dt} & = & -\frac{\partial h}{\partial q_1},
\eeq
with $h=q_1^3$. Solving these simple equations, we obtain:
\be
q_1(t) = q_1(0), \ \ \ p_1(t) = p_1(0) - 3 q_1(0)^2 t,
\ee
where $q_1(0)$ and $p_1(0)$ denote the values of $q_1$ and $p_1$
at time $t=0$. To obtain the action of the map
$\exp(:q_1^3:)$ on the phase space variables,
set $t=-1$ in the above equations and denote $q_1(-1)$, $p_1(-1)$
by $q_1^{fin}$, $p_1^{fin}$ and $q_1(0)$, $p_1(-0)$
by $q_1^{in}$, $p_1^{in}$ respectively.  Thus we get
\be
q_1^{fin} = q_1^{in}, \ \ \ p_1^{fin} = p_1^{in} + 3 (q_1^{in})^2.
\ee
Using Eq. (\ref{1b}), we can easily verify that the above result is
indeed correct.

Using the above procedure, we can identify symplectic maps $\exp(:h(z):)$
which give rise to polynomial mappings of the phase space variables
into themselves. These results \cite{grprep} can be codified
as the following simple principles which are easily generalized to
higher dimensions also.
\begin{enumerate}
\item{} All polynomials of the form $h(z)$ where both
a phase space variable and its
canonically conjugate variable \cite{Goldstein} do not occur simultaneously
give rise to polynomial symplectic maps via $\exp(:h(z):)$.
\item{} If a canonically conjugate pair $\{ q_i,p_i \}$ is present in the polynomial
$h(z)$, it only appears either in the form $a(z') q_i+g(p_i,z')$ or
$a(z') p_i + g(q_i,z')$
or its integer powers. Here $z'$ denotes the collection of phase space
variables $\{ q_j,p_k \}$ with $j \ne k \ne i$. Further, $a$ and $g$ are polynomials
in the indicated variables.
\end{enumerate}

We now return to the problem of symplectic integration. For the present,
we restrict ourselves
to one-turn symplectic maps in a two dimensional phase space truncated at order 4.
The results obtained below can be generalized to higher orders using symbolic
manipulation programs. The Dragt-Finn factorization of the symplectic map
is given by:
\be
{\cal M} = {\hat M} e^{:f_3:}\,e^{:f_4:},
\ee
where
\beq\label{fs}
f_3 & = & a_1 q_1^3 + a_2 q_1^2 p_1 + a_3 q_1 p_1^2 + a_4 p_1^3, \nonumber \\
f_4 & = & a_5 q_1^4 + a_6 q_1^3 p_1 + a_7 q_1^2 p_1^2 + a_8 q_1 p_1^3 + a_9 p_1^4.
\eeq
Here the coefficients $a_1, \ldots , a_9$ can be explicitly computed given
a Hamiltonian system\cite{Dragt}.
The above map captures the leading order nonlinearities of the system.
Since the action of the linear part ${\hat M}$ on
phase space variables is well known [cf. Eq. (\ref{3})] and is already a polynomial
action, we only refactorize the nonlinear part of map using polynomial maps.
It turns out that we require 7 polynomial maps for this purpose:
\be
{\cal M} \approx {\cal P} = {\hat M} e^{:h_1:}\,e^{:h_2:} \cdots e^{:h_{7}:},
\ee
where the numeral appearing in the subscript indexes the polynomial maps.
The $h_i$'s are given as follows:
\beq \label{polymap2d}
h_1 & = & b_1 q_1^3 + b_5 q_1^4, \ \ \
h_2  =  b_4 p_1^3 + b_9 p_1^4, \nonumber \\
h_3 & = & (b_2+b_3)(q_1 + p_1)^3, \ \ \
h_4  =  (b_3-b_2)(q_1-p_1)^3, \\
h_5 & = & (q_1+p_1+b_8 p_1^2)^3, \ \ \
h_6  =  (-q_1-p_1+b_6 q_1^2)^3, \ \ \
h_7  =  b_7 (q_1+p_1)^4. \nonumber
\eeq
Here $b_i$'s are at present unknown coefficients.
By forcing the refactorized form ${\cal P}$ to equal the original map ${\cal M}$
up to order 4 and using the CBH
theorem\cite{Cornwell}, we can easily compute these unknown coefficients in terms of
the known $a_i$'s. These expressions are given in the Appendix.
The explicit actions of the polynomial maps on the phase space variables
are also given there. This completely
determines the refactorized map ${\cal P}$. Each $\exp(:h_i:)$ is a
polynomial map which can be evaluated exactly and is explicitly symplectic.
Thus by using ${\cal P}$ instead of ${\cal M}$ in Eq. (\ref{rec}), we obtain
an explicitly symplectic integration algorithm. Further, it is fast to
evaluate and does not introduce spurious poles and branch points.
The above factorization is not unique. However, the principles outlined
earlier impose restrictions on the possible forms and this eases
considerably the task of refactorization. Moreover, we require the
coefficients $b_i$ to be polynomials in the known coefficients $a_i$.
Otherwise this can lead to divergences when $a_i$'s take on certain
special values. Finally, we minimize the number of polynomial maps
in the refactorized form. Our studies show that different polynomial
map refactorizations obeying the above restrictions do not lead to
any significant differences in their behavior.

The above refactorization has also been extended to
symplectic maps in a six dimensional phase space truncated at order 4.
In this case, we require 23 polynomial maps in the refactorization
to make ${\cal P}$ equal to ${\cal M}$ up to order 4:
\be
{\cal M} \approx {\cal P} = {\hat M} e^{:h_1:}\,e^{:h_2:} \cdots e^{:h_{23}:},
\ee
where the numeral appearing in the subscript indexes the polynomial maps.
Since listing out the explicit forms of the $h_i$'s and their coefficients
is not particularly illuminating, we do not list them here. However,
a FORTRAN program which implements the above polynomial map refactorization
is available from the author upon request.

We now analyze the leading order error committed in our method.
In our method, we first truncate the symplectic map to a given order
and then refactorize it using a product of polynomial maps. Both these
stages give rise to errors. When we truncate the symplectic map
${\cal M}$ at the $n$th order, we obtain
\be
{\cal M}_n = \hat M \exp(:f_{3}:)\,\,\exp(:f_{4}:)\, \ldots \exp(:f_{n}:).
\ee
The leading term that has been omitted is $\exp(:f_{n+1}:)$. From
properties of Lie transformations and Lie operators\cite{Dragt}, we have
\be
\exp(:f_{n+1}:) z = z+[f_{n+1},z]+\cdots \ ,
\ee
where $[,]$ denotes the usual Poisson bracket. Now, $[f_{n+1},z]$ gives
terms of the form $z^n$\cite{Dragt}. Thus error due to truncation of the
symplectic map is of order $z^n$.

Next, we refactorize the truncated symplectic map ${\cal M}_n$ as a
product of $k$ polynomial maps:
\be
{\cal M}_n = \hat M \exp(:h_{1}:)\,\,\exp(:h_{2}:)\, \ldots \exp(:h_{k}:).
\ee
These polynomial maps are obtained by first using the CBH series to
combine the Lie transformations
and then comparing with the original symplectic map. Both these
maps are made to agree up to order $n$. Therefore, the leading
error term is again of the form $\exp(:f_{n+1}:)$ giving rise
to an error of order $z^n$.

\section{Applications}

We now consider two applications of the above method.
The first example is to find the region of stability of
the following simple symplectic map:
\be
{\cal M} = {\hat M} \exp[:(q_1+p_1)^3:],
\ee
where
\be
{\hat M} = \pmatrix{\cos\theta&\sin\theta\cr-\sin\theta&\cos\theta },
\ee
and $\theta = \frac{\pi}{3}$.
We chose this example since the exact action of the
above map is known and hence the exact region of stability
can also be determined.
We found excellent agreement
between results obtained using polynomial maps and
the exact results.

We have also applied the method to more complicated Hamiltonian systems like
particle storage rings. We studied an electron storage ring with radio
frequency bunching cavities. The storage ring is composed of drifts, bending
magnets, quadrupoles, sextupoles and RF cavities. The efficacy of our method
is best revealed for such complicated
Hamiltonian systems. Since there are
many constituent elements (in storage rings like the Large Hadron Collider, there
can be thousands of elements), numerical integration using Hamiltonians for each
element is cumbersome and slow. On the other hand, a map based approach where
one represents the entire storage ring in terms of a single map is much faster \cite{yan}.
When this is combined with our polynomial map refactorization, one obtains a
symplectic integration algorithm which is both fast and accurate and is
ideally suited for such complex real life systems. The $q_3 - p_3$
phase plot for one million turns around the ring using our polynomial map method
is given in Figure 1. In this case, $q_3$ and $p_3$ represent the deviations
from the closed orbit time of flight and energy respectively. From theoretical
considerations, we expect the so-called synchrotron oscillations in these
variables. This manifests itself as ellipses in the phase space plot
of $q_3$ and $p_3$ variables (just as the oscillations of the simple
pendulum manifest themselves as ellipses in the coordinate-momentum phase
space plot). In Figure 1, we observe the expected
synchrotron oscillations.

\section{Conclusions}

To conclude, we have proposed a new symplectic integration algorithm based
on polynomial map refactorization. This should be of help in studying long
term stability of complicated accelerator systems and other Hamiltonian
systems.

\section*{Acknowledgements}

This work was supported by grants from DRDO and ISRO, India. The author is
also associated with the Jawaharlal Nehru Center for Advanced
Scientific Research as a honorary faculty member. The author would
like to thank the referees for their useful comments.

\section*{Appendix}

The coefficients $b_i$ in Eq. (\ref{polymap2d}) can be easily
determined using the CBH theorem\cite{Cornwell}. Their expressions
in terms of the known coefficients $a_i$ of the symplectic
map ${\cal M}$ [cf. Eq. (\ref{fs})] is given as follows:
\beq
b_1 & = & a_1-a_3/3, \ \ \
b_2  =  a_2/6, \nonumber \\
b_3 & = & a_3/6, \ \ \
b_4  =  a_4-a_2/3, \nonumber \\
b_6 & = & (2 a_6 -2 a_7 + a_8 + 18 b_1 b_2 - 36 b_2^2-36 b_1 b_3 + 36 b_3^2
+ 9 b_1 b_4 + 18 b_2 b_4 -18 b_3 b_4)/6, \nonumber \\
b_7 & = & (-a_6+2 a_7-a_8-18 b_1 b_2 + 36 b_2^2+18 b_1 b_3 - 36 b_3^2-9 b_1 b_4
-18 b_2 b_4 + 18 b_3 b_4)/4, \\
b_8 & = & (a_6 -2 a_7 + 2 a_8 + 18 b_1 b_2 - 36 b_2^2-18 b_1 b_3 + 36 b_3^2
+ 9 b_1 b_4 + 36 b_2 b_4 -18 b_3 b_4)/6, \nonumber \\
b_5 & = & a_5 - 9 b_1 b_2 - 9 b_2^2 + 9 b_3^2 - 3 b_6-b_7, \nonumber \\
b_9 & = & a_9-9 b_2^2 + 9 b_3^2 + 9 b_3 b_4 - b_7 - 3 b_8. \nonumber
\eeq
Note that the formulas have been sequenced in such a way that once
a given $b_i$ is evaluated, it is used in the formulas for the
$b_i$'s following it.

The actions of the polynomial maps $\exp(:h_i:)$ ($i=1,2, \ldots ,7$)
on the phase space
variables $q_1$, $p_1$ are easily evaluated using the procedure
outlined in the main text (see the discussion before Eq. (\ref{Heqns}).
We obtain the following results:
\beq
e^{:h_1:} q_1 & = & q_1, \ \ \
e^{:h_1:} p_1 = p_1 + 3 b_1 q_1^2 + 4 b_5 q_1^3, \nonumber \\
e^{:h_2:} q_1 & = & q_1 - 3 b_4 p_1^2 - 4 b_9 p_1^3 , \ \ \
e^{:h_2:} p_1 = p_1, \nonumber \\
e^{:h_3:} q_1 & = & q_1 - 3(b_2+b_3) (q_1+p_1)^2, \ \ \
e^{:h_3:} p_1  =  p_1 + 3(b_2+b_3) (q_1+p_1)^2, \nonumber \\
e^{:h_4:} q_1 & = & q_1 + 3(b_3-b_2) (q_1-p_1)^2, \ \ \
e^{:h_4:} p_1  =  p_1 + 3(b_3-b_2) (q_1-p_1)^2,  \\
e^{:h_5:} q_1 & = & q_1 - c_1 (1+2 b_8 p_1 + b_8 c_1), \ \ \
e^{:h_5:} p_1  =  p_1 + c_1, \nonumber \\
e^{:h_6:} q_1 & = & q_1 + c_2 , \ \ \
e^{:h_6:} p_1  =  p_1 - c_2 (1-2 b_6 q_1 - b_6 c_2), \nonumber \\
e^{:h_7:} q_1 & = & q_1 - 4 b_7 (q_1 + p_1)^3, \ \ \
e^{:h_7:} p_1  =  p_1 + 4 b_7 (q_1 + p_1)^3, \nonumber
\eeq
where
\be
c_1 = 3 (q_1 + p_1 + b_8 p_1^2)^2, \ \ \ c_2 = 3 (-q_1 - p_1 + b_6 p_1^2)^2.
\ee


\newpage

\section*{Figure Legends}

\begin{description}

\item{\bf Figure 1:} This figure shows the $q_3 - p_3$ phase
space plot for one million turns around a storage ring using
the polynomial map method (only every 1000th point is plotted).

\end{description}


\begin{references}
\bibitem[*]{byline} Electronic address: rangaraj@math.iisc.ernet.in
\bibitem{BNL} Stability of Particle Motion in Storage Rings, AIP
Conf. Proc. No. 292, eds. M. Month, A. Ruggiero and W. Weng (AIP,
New York, 1994)and references therein.
\bibitem{Dragt} A. J. Dragt,
in Physics of High Energy Particle Accelerators,
edited by R. A. Carrigan, F. R. Huson, and M. Month,
AIP Conference Proceedings No. 87 (American Institute of Physics,
New York, 1982), p. 147; A. J. Dragt, F. Neri, G. Rangarajan,
D. R. Douglas, L. M. Healy and R. D. Ryne, Ann.
Rev. Nucl. Part. Sci. 38 (1988) 455 and references therein.
\bibitem{channel} P. J. Channell and F. Neri, Fields Institute
Communications 10 (1996) 45.
\bibitem{list}R. D. Ruth, IEEE Trans. Nucl. Sci. 30 (1983) 2669;
Feng Kang, Proc. 1984 Beijing Symposium on Differential
Geometry and Differential Equations (ed. Feng Kang) (Science Press,
Beijing, 1985), p. 42;
F. Neri, Department of Physics Technical Report, University of Maryland (1988);
J. Irwin, SSC Report No. 228 (1989);
P. J. Channel and C. Scovel, Nonlinearity 3 (1990) 231;
H. Yoshida, Phys. Lett. A 150 (1990) 262;
E. Forest and R. Ruth, Physica D 43 (1990) 105;
G. Rangarajan, Ph. D. thesis, University of Maryland (1990);
G. Rangarajan, A. J. Dragt and F. Neri, Part. Accel.
28 (1990) 119;
A. J. Dragt, I. M. Gjaja, and G. Rangarajan, Proc. 1991 IEEE Part. Accel.
Conf.,  (1991) 1621;
A. J. Dragt and D. T. Abell, Int. J. Mod. Phys. A
(Proc. Suppl.) 2B (1993) 1019;
I. Gjaja, Part. Accel. 43 (1994) 133;
G. Rangarajan, J. Math. Phys. 37 (1996) 4514; for
associated group theoretical material see G. Rangarajan, J. Math. Phys. 38 (1997)
2710; G. Rangarajan, Pramana 48 (1997) 129;
G. Rangarajan, J. Phys. A: Math. and Gen. 31 (1998) 3649;
G. Rangarajan and M. Sachidanand, J. Phys. A: Math. and Gen. 33 (2000) 131.
\bibitem{yan} A. Chao, T. Sen, Y. Yan and E. Forest SSCL Report No. 459 (1991);
J. S. Berg {\em et. al.}, Phys. Rev. E 49 (1994) 722.
\bibitem{dragtprl} A. J. Dragt, Phys. Rev. Lett. 75 (1995) 1946.
\bibitem{irwin} J. Irwin, SSC Report No. 228 (1989).
\bibitem{list2} G. Rangarajan, Ph. D. thesis, University of Maryland (1990);
A. J. Dragt and D. T. Abell, Int. J. Mod. Phys. A
(Proc. Suppl.) 2B (1993) 1019;
G. Rangarajan, J. Math. Phys. 37 (1996) 4514.
\bibitem{etienne} E. Forest et. al. in Stability of Particle Motion in Storage Rings, AIP
Conf. Proc. No. 292, eds. M. Month, A. Ruggiero and W. Weng (AIP,
New York, 1994).
\bibitem{grsolve} G. Rangarajan, Ph. D. thesis, University of Maryland (1990);
G. Rangarajan, A. J. Dragt and F. Neri, Part. Accel.
28 (1990) 119;
G. Rangarajan and M. Sachidanand, J. Phys. A: Math. and Gen. 33 (2000) 131.
\bibitem{gjaja} I. Gjaja, Part. Accel. 43 (1994) 133.
\bibitem{Finn} A. J. Dragt and J. M. Finn, J. Math. Phys. 17 (1976)
2215.
\bibitem{Cornwell} J. F. Cornwell, Group Theory in Physics, volume
2 (Academic Press, London, 1984).
\bibitem{grprep} G. Rangarajan, manuscript in preparation (2001).
\bibitem{Goldstein} H. Goldstein, Classical Mechanics, 2nd ed. (Addison-Wesley,
Reading, 1980).

\end{references}
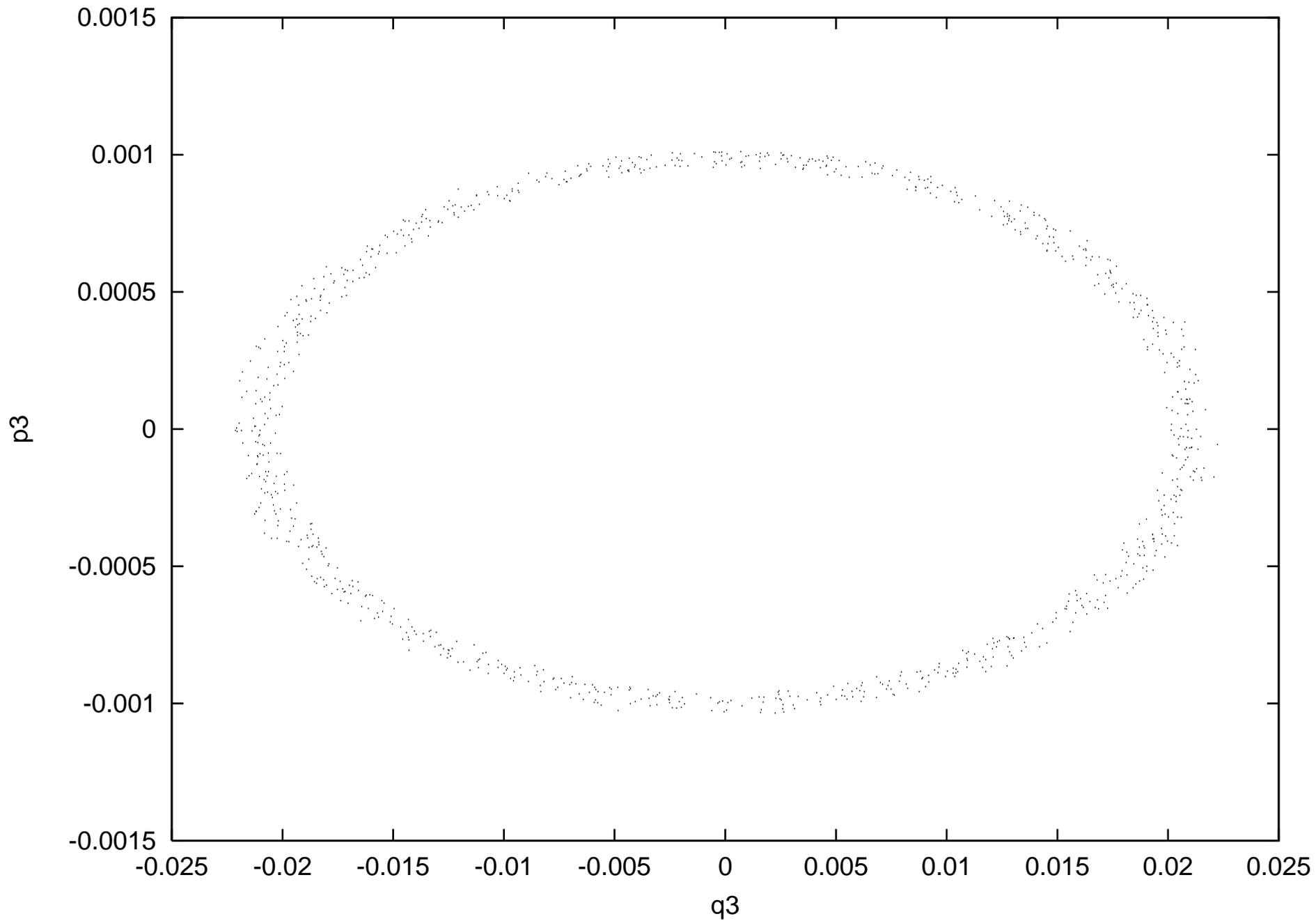
\end{document}